\documentclass[twocolumn, amsmath,amssymb,aps, prx,
showkeys,showpacs, floatfix] {revtex4-2}
\usepackage{graphicx}
\usepackage[usenames,dvipsnames]{color}
\usepackage{bm}
\usepackage[colorlinks=true,breaklinks=true,allcolors=blue]{hyperref}

\def\ri{{\rm i}}

\begin{document}

\title{Efficient Quantum Circuits based on the Quantum Natural Gradient}
\author{Ananda Roy}
\email{ananda.roy@physics.rutgers.edu}
\affiliation{Department of Physics and Astronomy, Rutgers University, Piscataway, NJ 08854-8019 USA}
\author{Sameer Erramilli}
\affiliation{Department of Physics and Astronomy, Rutgers University, Piscataway, NJ 08854-8019 USA}
\author{Robert M. Konik}
\affiliation{Division of Condensed Matter Physics and Material Science, Brookhaven National Laboratory, Upton, NY 11973-5000, USA}
\begin{abstract}
Efficient preparation of arbitrary entangled quantum states is crucial for quantum computation. This is particularly important for noisy intermediate scale quantum simulators relying on variational hybrid quantum-classical algorithms. To that end, we propose symmetry-conserving modified quantum approximate optimization algorithm~(SCom-QAOA) circuits. The depths of these circuits depend not only on the desired fidelity to the target state, but also on the amount of entanglement the state contains. The parameters of the SCom-QAOA circuits are optimized using the quantum natural gradient method based on the Fubini-Study metric. The SCom-QAOA circuit transforms an unentangled state into a ground state of a gapped one-dimensional Hamiltonian with a circuit-depth that depends not on the system-size, but rather on the finite correlation length. In contrast, the circuit depth grows proportionally to the system size for preparing low-lying states of critical one-dimensional systems. Even in the latter case, SCom-QAOA circuits with depth less than the system-size were sufficient to generate states with fidelity in excess of 99\%, which is relevant for near-term applications.  The proposed scheme enlarges the set of the initial states accessible for variational quantum algorithms and widens the scope of investigation of non-equilibrium phenomena in quantum simulators. 
\end{abstract}

\maketitle

\section{Introduction}
\label{sec:intro}
Quantum computation can be used to investigate strongly interacting quantum many body problems that lie beyond the reach of the classical computing paradigm~\cite{Feynman_1982, Lloyd1997}. In this context, the power of a quantum computer can be attributed to its ability to efficiently store and manipulate generic highly entangled states~\footnote{Note that this does {\it not} imply that classical efficiently solvable models do not generate highly entangled states, examples include certain states obtained using Clifford operations on a many-qubit system.}. The latter, in contrast to states with low entanglement, cannot be efficiently represented using conventional tensor-network-based approaches~\cite{Hastings2007, Vidal2008, Schuch2008, Verstraete2008}. These highly-entangled states of $n$ qubits can, in principle, be prepared using a quantum circuit of~${\cal O}(2^n)$ depth~\cite{Bergholm2005, Plesch2011, Malvetti_2021}. Intuitively, this can be understood as a consequence of separately encoding the information associated with all the possible amplitudes in quantum gates~\footnote{Free-fermionic models are exceptions where an efficient scheme for creating all states in~${\cal O}(n\ln n)$ depth has been proposed~\cite{Verstraete2009}. See also Refs.~\cite{Fishman2015, Wu2022} for approximate approaches.}. The exponentially large circuit-depth can be reduced to~${\cal O}(n^2)$, but at a cost of increasing the circuit-width to~${\cal O}(2^n)$ qubits~\cite{Araujo_2021}. In the noisy intermediate scale quantum era~(NISQ), where quantum simulators typically host~$n\sim10^2$ qubits with modest coherence properties, it is desirable to do away with the exponential in both the circuit width and depth. To that end, several approaches have been suggested based on low-rank state-preparation~\cite{Araujo_2023} as well as protocols for creation of matrix product states on quantum circuits~\cite{Jobst2022, Cruz2022, Malz2023}. 

An alternative to the aforementioned is a variational approach based on hybrid quantum-classical optimization which lies at the heart of a large number of NISQ-era algorithms such as the variational quantum eigensolver~\cite{Peruzzo2011, Tilly2022} and the quantum approximate optimization algorithm~\cite{Farhi2014, Lloyd2018, Hadfield2019, Morales2020}. In this approach, a parameterized quantum circuit~(pQC) built out of a pool of unitary rotations is used as an ansatz to create the target quantum state. The parameters of this quantum circuit are subject to an optimization procedure that minimizes a cost function. The latter can be, for instance, the expectation value of a target Hamiltonian or negative overlap to the target state. While the application of the unitary rotation as well as measurement of the relevant cost function are implemented on a quantum simulator, the optimization is performed on a classical computer. The result of the classical optimization is subsequently used as an updated guess for the parameters of the quantum circuit. The process is repeated until the desired convergence criterion for the state preparation is reached. This variational approach has been successfully used to generate the ground states of a number of quantum Hamiltonians~\cite{Ho2019, Wierichs2020}. We note that success of such a variational algorithm is not guaranteed in general. The process of training the circuit can be NP-hard~\cite{Bittel_2021} and local minima and barren plateaus induced by noise/over-parameterization can severely degrade the performance of the algorithm. To mitigate some of the detrimental effects of the aforementioned, several approaches have been proposed, see, for example, Refs.~\cite{Wang_2021, Akshay_2021, wurtz2021, galda2021, Lee_2021, Farhi_2022, pelofske2023, Shaydulin_2023, Shaydulin_2024}. 

Naturally, the variational quantum-classical approach relies on:  i) the choice of the ansatz quantum circuit and ii) the efficacy of the classical optimization procedure. For a target ground state of Hamiltonian~$H_T$, the ansatz quantum circuit can be built of a fixed number~(say, $p$) of alternating applications of~${\rm exp}(-\ri\theta_jH_T)$ and~${\rm exp}(-\ri\phi_jH_M)$, where~$H_M$ is a mixing Hamiltonian~\cite{Farhi2014, Hadfield2019, Wang2020}. The~$2p$ parameters~$\{\theta_j, \phi_j\}$ are subject to the classical optimization procedure~\footnote{Note that Ref.~\cite{Ho2019} uses a different choice of $H_T, H_M$ for state-preparation.}. The latter can be any standard optimization routine such as a first-order gradient descent~\cite{Kingma2017}, a quasi-second order method like BFGS~\cite{Fletcher2013} or the quantum natural gradient~(QNG)~\cite{Stokes2020, Gacon2021}. In contrast to the first two, the QNG method finds an optimization path along the steepest gradient direction, but takes into account the geometry of the manifold of quantum states~\cite{Zanardi2007, Kolodrubetz2013, Kolodrubetz2017}. The QNG method has been shown to be more robust than the others for finding the optimal set of parameters for the quantum circuit~\cite{Wierichs2020} and is used in the current work. 

In this work, we propose a general symmetry-conserving quantum circuit ansatz based on the modification~\cite{Ho2019} of the well-known quantum approximate optimization algorithm~(QAOA) ansatz~\cite{Farhi2014}. The main characteristics of the SCom-QAOA circuits are summarized below. First, these circuits allow efficient preparation of arbitrary quantum states, including non-translation-invariant states despite starting from a translation-invariant initial state. In fact, for translation-invariant states, our quantum circuit ansatz is similar to that in Ref.~\cite{Ho2019}. Second, the pQC ansatz is chosen while conserving the symmetries of the model, when possible. This restricts the pool of allowed unitary rotations. Third, the depth of the circuit is determined by not only the desired fidelity, but also the amount of entanglement contained in the target state. In this work, we focus exclusively on pure target states and thus, von Neumann entropy of a subsystem serves as a useful quantitative measure of the amount of the entanglement. For a given desired fidelity, ground states of gapped one-dimensional Hamiltonians are generated by a circuit whose depth that depends not on the system-size, but rather on the correlation-length of the system. The bounded correlation length of these systems leads to a bounded entanglement entropy for the ground states of these systems~\cite{Hastings2007}, which in turn leads to a bounded circuit depth. The depth for realization of critical one-dimensional system scales with system-size. Fourth, the proposed approach generates eigenstates of lattice Hamiltonians that are integrable as well as non-integrable without any discernible difference in performance, indicating the robustness of this approach. This is shown by creating low-lying eigenstates of non-integrable perturbed Ising and tricritical Ising lattice models. Finally, the proposed scheme can be directly implemented for creation of eigenstates of generic two-dimensional quantum Hamiltonians - a key set of problems where NISQ simulators have the potential to outperform classical computers. We note that several variants of the QAOA ansatz have been proposed, see Refs.~\cite{Wang_2020, LaRose_2022, Fuchs_2022, He_2023, Golden_2023}. In contrast to the previous approaches, the SCom-QAOA ansatz is built entirely out of the target Hamiltonian whose eigenstates are being computed. Importantly, the ansatz is constructed while respecting the symmetries that are conserved by the target Hamiltonian. This reduces the space where the search for the target space occurs, increasing the efficacy of the proposed ansatz. 

The proposed pQC ansatz is demonstrated by maximizing the square of the overlap to a target state obtained independently using density matrix renormalization group~(DMRG) technique on matrix product states. The evolution of the quantum circuit is performed using the time-evolved block-decimation~(TEBD) algorithm. As such, the simulations performed use perfect circuits and qubits without accounting for finite gate-errors or lifetimes of the qubits. This can be remedied by using a noisy quantum circuit simulator such as Qiskit~\cite{Qiskit}. In fact, as shown in Ref.~\cite{Lamb2024}, despite the presence of a larger parameter space, it is feasible to perform the necessary optimization with circuit depths that are reasonable in the NISQ era. 

The proposed QNG optimization of the SCom-QAOA circuits is a modified version of the general problem of determining quantum circuit complexity~\cite{Nielsen2005, Nielsen2006, Dowling2006} in terms of a geometric control theory~\cite{Jurdjevic1997}. For the general case, the relevant unitary operator that takes the initial state to a given target state can be shown to be approximated by~${\cal O}(n^6 D^3)$ single and two-qubit gates~\cite{Nielsen2006}. Here,~$D$ is the distance between the identity and the desired unitary operator~\cite{Nielsen2005}. In this work, the optimization is performed while first fixing the set of unitary rotations that are used in the pQC and subjecting the angles of these unitary rotations to the optimization routine. Therefore, the optimization occurs on a sub-manifold of the entire space of unitary rotations. A broader optimization in the space of unitary rotations, as proposed in Ref.~\cite{Nielsen2006}, is likely to yield more versatile quantum circuits. In fact, some progress has already been achieved to efficiently perform this optimization~\cite{Luchnikov2021}. 

The article is organized as follows. In Sec.~\ref{sec:QNG}, the QNG-based optimization scheme is described, including details on the pQC ansatz and the necessary details about the Fubini-Study metric. Sec.~\ref{sec:res_1D} presents results of numerical simulations for the perturbed Ising and tricritical Ising models. Sec.~\ref{sec:concl} provides a concluding summary and outlook.

\section{QNG Optimization Scheme for SCom-QAOA Circuits}
\label{sec:QNG}
Here, the details of the ansatz for the quantum circuit and the QNG optimization method are presented. The case of open boundary conditions is described below. The periodic case can be analyzed  analogously. 

\subsection{SCom-QAOA Quantum Circuits}
Consider the general problem of preparation of arbitrary eigenstates of target quantum Hamiltonians that can be decomposed as:
\begin{equation}
\label{eq:H}
H_T = H_{\rm ons} + H_{\rm nn} + H_{\rm nnn} + \ldots,
\end{equation}
where ons, nn, nnn denote the terms of the Hamiltonian that involve one site, nearest-neighbors, next-nearest-neighbors and so on. Each of the aforementioned kinds of interaction terms can be divided into groups, labeled by~$s_\alpha$,~$\alpha = {\rm ons}$, nn, nnn \ldots:
\begin{align}
\label{eq:H_dec}
H_{\alpha} &= \sum_{s_\alpha}K^{s_\alpha} =\sum_{s_\alpha}\sum_{j}\lambda^{s_\alpha}_jK^{s_\alpha}_j,
\end{align}
where~$K^{s_\alpha}_j$ is the corresponding operator in the Hamiltonian and~$\lambda^{s_\alpha}_j$ is the coupling at the~$j^{\rm th}$-site.
For example, the Hamiltonian for the one-dimensional transverse field Ising model in the presence of a longitudinal field is 
\begin{align}
\label{eq:H_I}
H^{\rm I} = -\lambda^X\sum_{j = 1}^LX_j - \lambda^Z\sum_{j = 1}^LZ_j -\sum_{j = 1}^{L-1}X_{j}X_{j+1},
\end{align}
where~$\lambda^{X,Z}$ are strengths of the longitudinal and transverse fields respectively. Then, the decomposition of Eq.~\eqref{eq:H_dec} yields~$s_{\rm ons}\in\{X,Z\}$ and~$s_{\rm nn}\in\{XX\}$. Then,
\begin{equation}
K^X_j = - X_j, K^Z_j = -Z_j, K^{XX}_j = -X_jX_{j+1}. 
\end{equation}

While for the Ising case, the grouping described in Eq.~\eqref{eq:H_dec} is unique, this is not true in general. For instance, consider the XXZ spin-chain Hamiltonian:
\begin{align}
\label{eq:H_XXZ}
H^{\rm xxz} &= -\sum_{j = 1}^{L-1}\left[X_jX_{j+1} + Y_jY_{j+1} + (\cos\gamma) Z_jZ_{j+1}\right],
\end{align}
where~$\gamma$ is the anisotropy parameter. One possible decomposition would be~$s_{\rm nn}\in\{XX, YY, ZZ\}$ with
\begin{equation}
\label{eq:H_xxz_dec_1}
K^{AA}_j = -A_jA_{j+1},\ A = X, Y, Z.
\end{equation}
An equally valid decomposition would be with~$s_{\rm nn}\in\{XX+YY, ZZ\}$ with 
\begin{align}
\label{eq:H_xxz_dec_2}
K_j^{XX + YY} &= -(X_jX_{j+1} + Y_jY_{j+1})\nonumber,\\ K_j^{ZZ} &= -Z_jZ_{j+1}.
\end{align}
The two groupings for the XXZ chain lead to two different SCom-QAOA circuits that conserve different symmetries of the XXZ model. This is because the operators~$K_j^{s_\alpha}$-s will be used to generate the unitary rotations of the different layers of the SCom-QAOA circuit. The relevant unitary operators and the corresponding symmetries are described next. 

\begin{figure}\centering
\includegraphics[width = 0.5\textwidth]{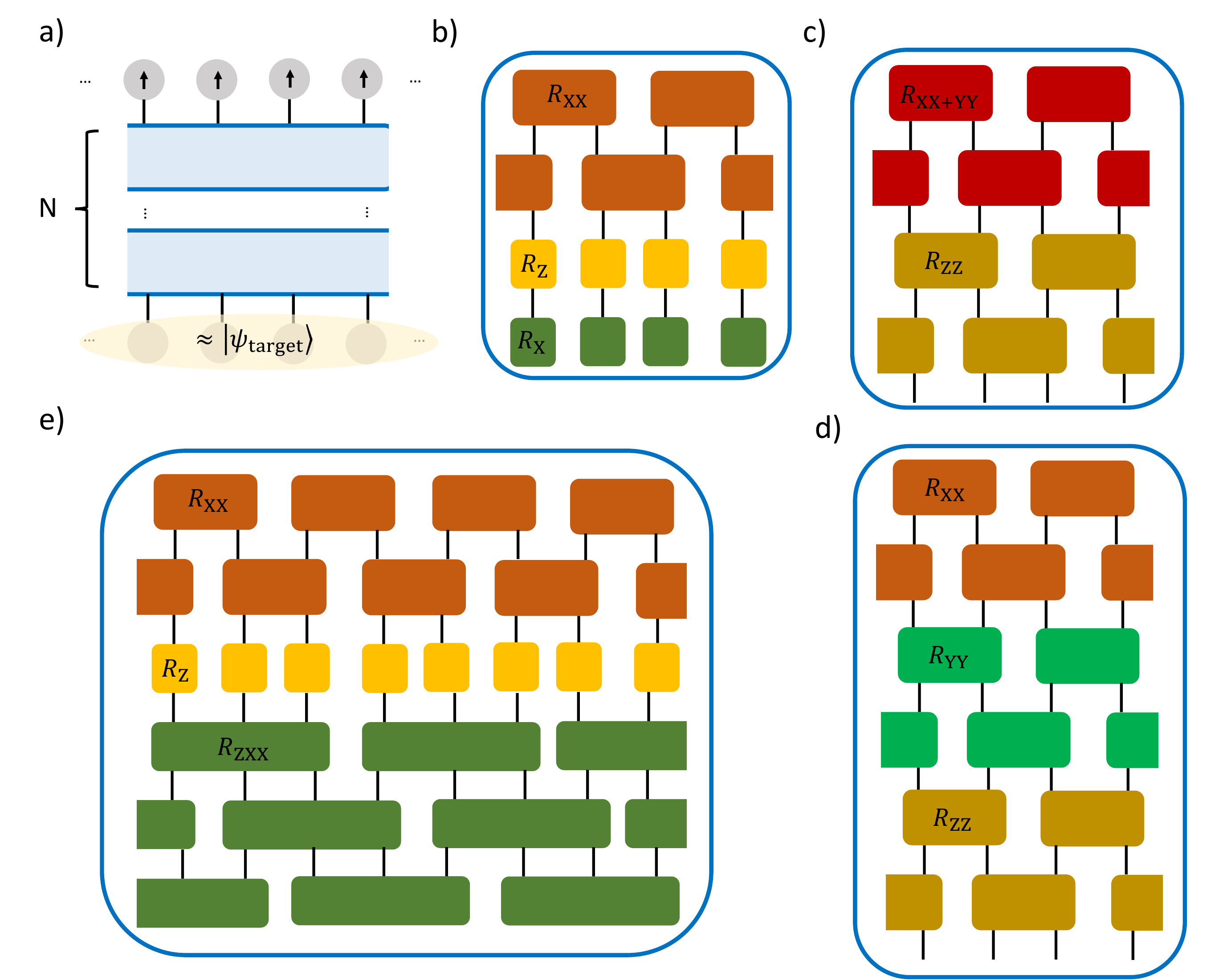}
\caption{\label{fig:schematic} (a) Scheme for SCom-QAOA circuits. For definiteness, the initial state is shown to be the product state~$|\uparrow\rangle^{\otimes L}$. The blue boxes represent the unitary rotations at different layers~[Eqs.~(\ref{eq:U}, \ref{eq:U_l})]. The generators of the unitary rotations do not change from layer to layer, only the angles.~(b) The unitary operator,~$U_l$, for the Ising model with the longitudinal field~[Eq.~\eqref{eq:H_I}]. Note that the angles for the different unitary operators are, in general, site-dependent~[Eq.~\eqref{eq:U_subl}].~(c,d) The unitary operator,~$U_l$ for the XXZ spin-chain. Note that while the target state can be reached using either unitary, the symmetry conserved by the quantum circuit is different~[Eqs.~(\ref{eq:H_xxz_dec_1}, \ref{eq:H_xxz_dec_2})].~(e) The unitary operator,~$U_l$ for the spin-chain Hamiltonian of Eq.~\eqref{eq:H_TCI}, which for certain choices of parameters, gives rise to the tricritical Ising model~(see Sec.~\ref{sec:TCI} for more details). This quantum circuit conserves the~$\mathbb{Z}_2$ charge conserved by the Hamiltonian of Eq.~\eqref{eq:H_TCI}. Note that in panels~(c,e), only the leading order terms of~$\tilde{U}^l_\alpha$~[Eq.~\eqref{eq:U_subl_g1}] are shown for the sake of brevity.}
\end{figure}

The pQC ansatz that performs the desired unitary rotation from the initial state to the target state is given by
\begin{equation}
\label{eq:U}
U = \prod_{l = 1}^{N}U^l,
\end{equation}
where~$N$ the total number of layers,~$U^l$ is the unitary rotation at the~$l^{\rm th}$ layer and it is understood that~$l = 1$ acts first. The unitary~$U^l$ is decomposed further into unitary rotations of sublayers as:
\begin{align}
\label{eq:U_l}
U^l = \ldots U^l_{\rm nnn}U^l_{\rm ons}U^l_{\rm nn},
\end{align}
where the dots indicate unitary operators generated by Hamiltonian terms which are beyond next-nearest neighbor. The unitary operator~$U^l_\alpha$ for the sublayer~$\alpha$ is defined as:
\begin{equation}
\label{eq:U_subl}
U^l_\alpha = {\rm exp}\big({-\ri \sum_{s_\alpha}\sum_j\theta^{s_\alpha}_{l,j}K^{s_\alpha}_j}\big),
\end{equation}
where~$\alpha = {\rm ons}$, nn, nnn, \ldots and~$s_\alpha, K^{s_\alpha}_j$-s have been defined in Eq.~\eqref{eq:H_dec}. The angles~$\theta^{s_\alpha}_{l,j}$ are the parameters that are subject to the QNG-optimization process. Note that although the angles~$\{\theta_{l,j}^{s_\alpha}\}$ change from layer to layer, the operators~$K^{s_\alpha}_j$ generating the unitary rotation remain the same~\footnote{This assumption can be relaxed in generalizations of the pQC ansatz proposed in this work.}. The operator~$U^l_\alpha$ is decomposed into product over unitary operators corresponding to the different groupings:
\begin{equation}
\label{eq:U_subl_g1}
\tilde{U}^l_\alpha = \prod_{s_\alpha}{\rm exp}\big({-\ri \sum_j\theta^{s_\alpha}_{l,j}K^{s_\alpha}_j}\big).
\end{equation}
Clearly, there are different possible~$\tilde{U}^l_\alpha$-s starting with the same~$U^l_\alpha$ since the different terms in the summation over~$s_\alpha$ in Eq.~\eqref{eq:U_subl} do not necessarily commute with each other. In the models analyzed in this work, the different possible choices led to minor quantitative changes in the results. For instance, in the case of the Ising Hamiltonian~[Eq.~\eqref{eq:H_I}], the order of the~$R_X$ and~$R_Z$ unitary operators~[Fig.~\ref{fig:schematic}(b)] did not make any qualitative difference in the performance of the SCom-QAOA circuit. Finally, a further simplification arises when the terms at different sites within each group commute: 
\begin{equation}
\label{eq:K_comm}
\big[K_j^{s_\alpha},K_k^{s_\beta}\big] = 0, {\rm if\ } s_\alpha=s_\beta
\end{equation}
for all choices of~$j,k$. In this case, a further simplification occurs and the sublayer unitary operator reduces to:
\begin{equation}
\label{eq:U_subl_g2}
\tilde{U}^l_\alpha = \prod_{s_\alpha}\prod_{j}{\rm exp}\big({-\ri \theta^{s_\alpha}_{l,j}K^{s_\alpha}_j}\big).
\end{equation}
Evidently, Eq.~\eqref{eq:U_subl_g2} does not hold for all groupings of generic models. Consider, for example, for the XXZ chain. For the grouping of Eq.~\eqref{eq:H_xxz_dec_1}~[see Fig.~\ref{fig:schematic}(d)], Eq.~\eqref{eq:K_comm} is valid, but not for the decomposition of Eq.~\eqref{eq:H_xxz_dec_2}~[see Fig.~\ref{fig:schematic}(c) for the leading order contribution]. In the latter case, a higher-order Suzuki-Trotter decomposition of a suitable order can be used to implement Eq.~\eqref{eq:U_subl_g1}. In all cases considered in this work, the leading order term presented in Eq.~\eqref{eq:U_subl_g2} was sufficient to reach the desired fidelity. 

There are two important features of the ansatz pQC. First, the ansatz pQC can be chosen to conserve a given symmetry of a model. For the pQC to conserve the charge~$Q$, where~$[Q, H_T] = 0$, in general,
\begin{equation}
\bigg[Q, {\rm exp}\big({-\ri\sum_j \theta^{s_\alpha}_{l,j}K^{s_\alpha}_j}\big)\bigg] = 0, 
\end{equation}
which is satisfied for~$[Q, K^{s_\alpha}_j] = 0$ for each~$s_\alpha$. This restricts possible decompositions of Eq.~\eqref{eq:H_dec}. Consider the two possible decompositions of the XXZ chain Hamiltonian in Eqs.~(\ref{eq:H_xxz_dec_1}, \ref{eq:H_xxz_dec_2}). If the pQC has to conserve the U(1) symmetry associated with~$Q_1 = \sum_{j}Z_j$, only Eq.~\eqref{eq:H_xxz_dec_2} is permitted. On the other hand, for the pQC to conserve the~$\mathbb{Z}_2$-symmetry associated with~$Q_2 = \prod_jZ_j$, either choice is allowed.  Note that at the isotropic point, the XXZ chain has~$SU(2)$-symmetry. A SCom-QAOA circuit can be constructed while conserving this non-abelian symmetry as well. The simplest case, where all the~$\theta^{s_\alpha}_{l,j}$-s are site-independent, has been analyzed in Ref.~\cite{Ho2019}.  Second, the angles~$\theta_{l,j}^{s_{\alpha}}$ are chosen to be site-dependent. This is crucial to reach non-translation-invariant target states starting with translation-invariant initial states. However, if the goal is to find a translation-invariant eigenstate of a translation-invariant Hamiltonian, the SCom-QAOA circuit ansatz can be simplified and it is sufficient to choose~$\theta_{l,j}^{s_\alpha} = \theta_l^{s_\alpha}$. In this case, the number of circuit parameters to be optimized is equal to the number of layers times the number of sublayers. This is in contrast to the general case when the number of parameters is larger by a factor of the number of sites. The site-independent case leads to straightforward generalization of the ansatz of Ref.~\cite{Ho2019}, albeit with a different optimization procedure based on the QNG. The latter is explained below. 

\subsection{Quantum Natural Gradient Optimization}
The QNG-optimization procedure is described for the parameters~$\vec{\Theta}\equiv\{\theta_{l,j}^{s_\alpha}\}$. In contrast to conventional optimization routines like BFGS, the QNG ensures that optimization follows the steepest descent taking into account the geometry of the space of wavefunctions~\cite{Stokes2020, Zanardi2007} and can viewed as the quantum analog of approach based on information geometry~\cite{Amari1998}. In the quantum context, this has been considerably successful in preparing ground states of spin models such as the transverse field Ising model and the Heisenberg spin chain~\cite{Stokes2020, Wierichs2020}. However, the potential of this method for simulation of general eigenstates of lattice Hamiltonians that realize interacting quantum field theories in the scaling limit has not been investigated. This work generalizes the previous works to realize arbitrary excited states of interacting lattice models and in particular, demonstrates the qualitatively different behavior of the QNG-optimizer for gapped vs gapless systems. The relevant cost function~${\cal L}(\vec\Theta)$ to be minimized can be the expectation value of some target Hamiltonian~$H$, denoted by~${\cal L}_H$, or the negative of the square of the overlap to a target state~$\psi_T$, denoted by~${\cal L}_O$:
\begin{align}
\label{eq:L_H}
{\cal L}_H(\vec\Theta) &= \big\langle\psi(\vec\Theta)\big|H\big|\psi(\vec\Theta)\big\rangle, \\
\label{eq:L_O}
{\cal L}_O(\vec\Theta) &= -\big|\big\langle\psi_T\big|\psi(\vec\Theta)\big\rangle|^2.
\end{align}
The parameters~$\vec\Theta$ can be iteratively obtained from the equation~\cite{Stokes2020}:
\begin{equation}
\label{eq:theta_tp1}
\Theta_p(t+1) = \Theta_p(t) - \eta \sum_q g(\vec{\Theta})_{pq}^{-1}\cdot\frac{\partial{\cal L}(\vec{\Theta})}{\partial \Theta_q(t)}, 
\end{equation}
where~$t$ is the iteration index and~$\eta$ is the learning rate, typically a real number between 0 and 1. The indices~$p, q$ denote the indices of the reshaped vector~$\vec\Theta$. Finally, the matrix~$g(\vec\Theta)$ is the Fubini-Study metric tensor. The latter incorporates the geometry of the manifold of wavefunctions while computing the steepest gradient. This is then used during the iterations parameterized by~$\vec\Theta(t+1)$ and~$\vec\Theta(t)$~\footnote{Replacing the~$g$ by the identity matrix recovers the ordinary first order gradient descent.}. The Fubini-Study metric tensor is the real part of the quantum geometric tensor,~$G(\vec\Theta)$:
\begin{align}\label{eq:G_def}
G_{pq}(\vec{\Theta}) &= \Big\langle \frac{\partial\psi(\vec\Theta)}{\partial \Theta_p}\Big| \frac{\partial\psi(\vec\Theta)}{\partial \Theta_q}\Big\rangle\nonumber\\\qquad& - \Big\langle \frac{\partial\psi(\vec\Theta)}{\partial \Theta_p}\Big| \psi(\vec\Theta)\Big\rangle\Big\langle \psi(\vec\Theta)\Big| \frac{\partial\psi(\vec\Theta)}{\partial \Theta_q}\Big\rangle,\\
g_{pq}(\vec{\Theta}) &\equiv \Re \big[G_{pq}(\vec{\Theta})\big].
\end{align}
In practical computations, often the metric tensor~$g_{pq}$ is singular. In these cases, it is convenient to introduce a Tikhonov regularization parameter~$\delta$~\cite{Wierichs2020}:~$g\rightarrow g + \delta \mathbb{I}$. This regularized metric tensor can then be inverted in computing~$\vec\Theta$-s. Thus, the computation of~$G, {\cal L}$ corresponds to evaluation of multi-point correlation-functions of suitable operators for the states generated by the SCom-QAOA circuit. Explicitly, 
\begin{align}
\label{eq:G_def}
G_{pq} &= \big\langle \psi_p\big|K_p ({U_p^>})^\dagger U_q^> K_q\big|\psi_q\big\rangle\nonumber\\\qquad& - \big\langle\psi_p\big|K_p\big|\psi_p\big\rangle\big\langle\psi_q\big|K_q\big|\psi_q\big\rangle.
\end{align}
Here,~$|\psi_p\rangle$ is the state generated at the~$p^{\rm th}$ step. For instance, if~$\Theta_p = \theta_{l,j}^{s_\alpha}$, the state~$|\psi_p\rangle$ is generated by applying the quantum circuit up to the sublayer~$s_{\alpha}$ of the layer~$l$. The operator~$U_p^>$ stands for the remaining unitary rotations that evolve the state~$|\psi_p\rangle$ for the remaining layers and sublayers of the quantum circuit. It is easy to explicitly verify that Eq.~\eqref{eq:G_def} implies that the metric tensor is symmetric, as expected for a metric governing the distance between two quantum states:
\begin{align}
\label{eq:G_symm}
G = G^\dagger\Rightarrow g = g^{T},
\end{align}
where T indicates transposition. In actual computations, it is often advantageous to compute only the upper triangular and diagonal parts of~$G$, with the rest of the tensor obtained using Eq.~\eqref{eq:G_symm}. The gradient of the overlap and the Hamiltonian cost functions are given by:
\begin{align}
\frac{\partial{\cal L}_H}{\partial\Theta_p} &= 2\Re\big\langle \psi_p\big|iK_p(U_p^>)^\dagger H\big|\psi(\vec\Theta)\big\rangle,\nonumber\\
\frac{\partial{\cal L}_O}{\partial\Theta_p} &= -2\Re\Big[\big\langle\psi_p\big|iK_p(U_p^>)^\dagger\big|\psi_T\big\rangle\big\langle\psi_T\big|\psi_p\big\rangle\Big].
\end{align}
The computation of the metric tensor and the relevant cost functions can be straightforwardly implemented using a time-evolution algorithm present in a tensor network package. This is done here using the TEBD algorithm of the TeNPy package~\cite{DMRG_TeNPy}. 
 
\section{Results}
\label{sec:res_1D}
Here, it is shown that the QNG-based optimizer for the SCom-QAOA circuits enable generation of ground states of gapped 1D Hamiltonians with circuit-depth that is proportional to the correlation length, while the same is proportional to the system-size for the gapless case. The two cases are demonstrated by investigating the critical Ising and tricritical Ising chains as well as the magnetic perturbation of the Ising model. While the results are presented for open boundary conditions, similar results were obtained for periodic boundary conditions which are not shown for brevity.

For the numerical computations, the overlap cost-function~[Eq.~\eqref{eq:L_O}] was used since this was found to be more stable for targeting excited states of the different models, with the target state obtained using DMRG technique. For both the DMRG and the TEBD computations, the maximum bond-dimension was chosen to be always higher than what was reached during any of the TEBD evolution of the quantum circuit. The singular value cutoff was chosen to be~$10^{-10}$. This was to ensure that TEBD would capture faithfully the dynamics of a quantum simulator keeping singular values up to $10^{-12}$. The cut-off for the desired fidelity~(defined as the absolute value of the overlap) to the target state,~$|\psi_T\rangle$, is set to 99\% in this work. This is chosen since the typical two-qubit gate-error in current noisy simulators is of that order. Similar results were obtained for higher cut-offs. The learning rate~$\eta$ was set to 0.25 and a regularization parameter~$\epsilon=0.01$ was used while inverting the metric~$g$ in Eq.~\eqref{eq:theta_tp1}. The initial choice for the angles~$\theta_{l,j}^{s_\alpha}$ was taken to be~$= 0.01$, but the QNG-based optimization procedure was mostly insensitive to the initial choice of angles. Choosing too large a learning rate led to oscillator behavior while during the QNG-based optimization procedure. An `optimal' range of~$\eta$ was chosen based on trial and error. While we chose the learning rate to be constant for a given problem, using an~$\eta$ that depends on the number of iterations yielded similar results and could be used as well. Furthermore, for certain excited states, the QNG optimization got stuck in a local minima when starting from all angles being equal. In these cases, a random set of angles for~$\theta_{l,j}^{s_\alpha}$ was used. This can be further improved by the use of basin-hopping optimization or randomization of initial states, see also Refs.~\cite{Golden_2023b, Pelofske_2023, Golden_2023a}.  

The qubits of the circuit were initialized in the perfectly-polarized state~$|\uparrow\rangle^{\otimes L}$. Note that any other product state can be related to this state by circuit of unit depth and as such, modifies our results only by a depth of ~${\cal O}(1)$. Also, the so-called `block-diagonal approximation' of the Fubini-Study metric~$g$ was not successful in generating the many-body states investigated in this work. This is indicative of correlations between different sublayers and different layers in the quantum circuit evolution, which is not surprising and matches the findings in Ref.~\cite{Wierichs2020}. 

\subsection{The critical Ising chain}
\label{sec:crit_Ising}
Here, the results for the critical transverse field Ising model are shown. The relevant Hamiltonian is given in Eq.~\eqref{eq:H_I} with~$\lambda^Z =1$, $\lambda^X = 0$. The corresponding layer unitary~$U^l$~[Eq.~\eqref{eq:U_l}] is built out of a layer of $R_{\rm XX}$-s followed by a layer of~$R_{\rm Z}$-s~(Fig.~\ref{fig:schematic}, top right panel, without the layer of~$R_{\rm X}$ rotations). For circuit parameters targeting the~$Q =\prod_jZ_j= +1$ symmetry sector, the initial state was chosen to be:~$|\psi_0\rangle = |\uparrow\rangle^{\otimes L}$~($L$ equals the system-size). When searching for the parameters to realize states in the~$Q = -1$ sector, the initial state was chosen to be~$|\psi_0\rangle=X_j|\uparrow\rangle^{\otimes L}$, where~$j$ is an integer close to~$L/2$. 

\begin{figure}\centering
\includegraphics[width = 0.5\textwidth]{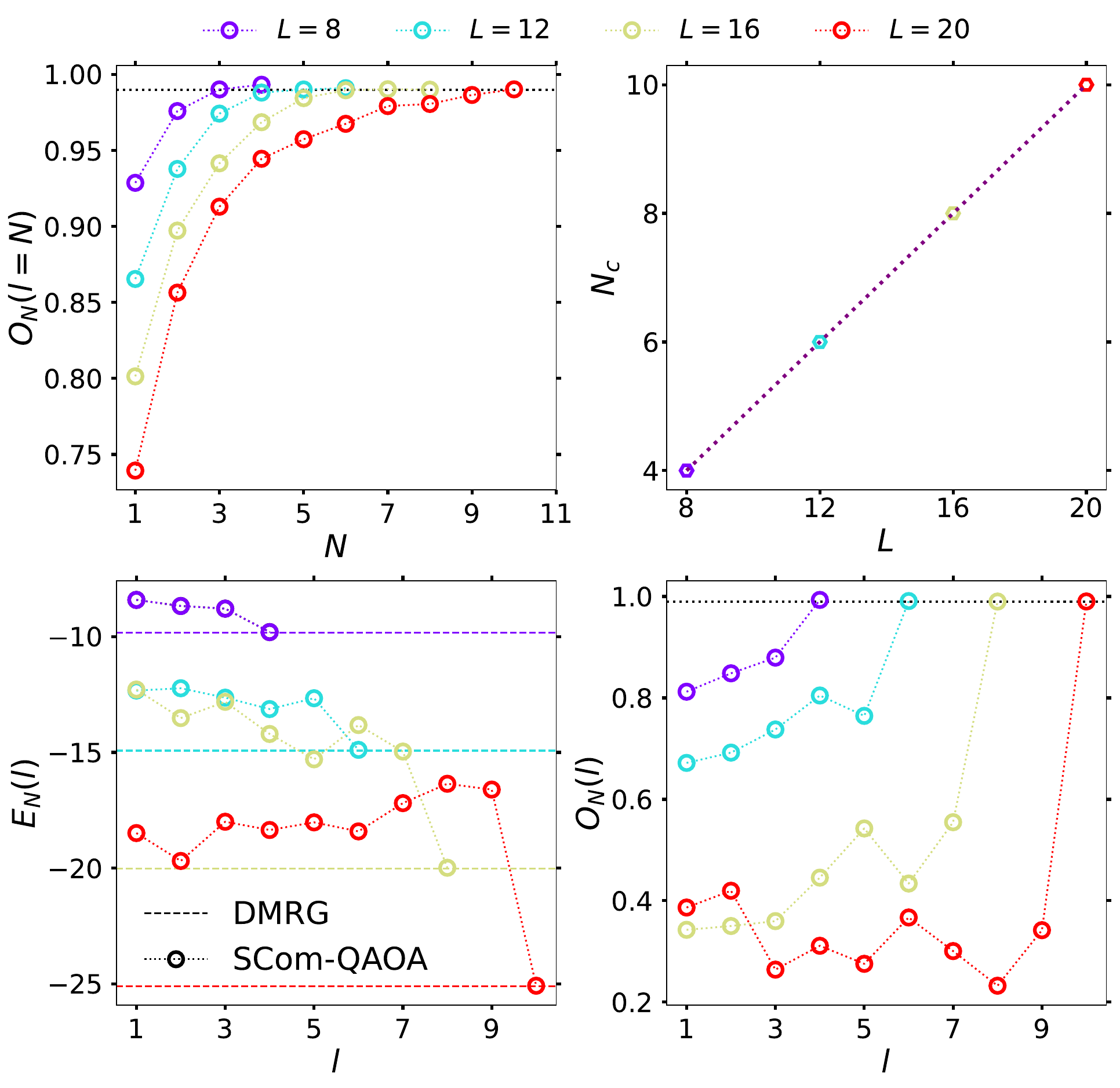}
\caption{\label{fig:TFI_gs} Results for the case when the target state is the ground state of Eq.~\eqref{eq:H_I}. The black dotted lines in the top left and bottom right panels correspond to the desired fidelity of 0.99. (Top left) Variation of the fidelity to the target state~$O_N(l = N) = |\langle \psi_T|U|\psi_0\rangle|$ at the end of the $N$-layers of the SCom-QAOA circuit. Here,~$l$ is the layer index. (Top right) The number of layers required to reach the desired fidelity,~$N_c$, was found to grow proportionally with the system size~$L$.~(Bottom left) Variation of the obtained energy with layer index~$l$ for~$N = N_c$. The dashed lines indicate the DMRG results for the target state.~(Bottom right) Variation of~$O_{N}(l)$ as the circuit evolves through the different layers for~$N=N_c$. In the two bottom panels, the variations of both overlap and energy is non-monotonic. This is compatible with the number of required layers to reach desired fidelity proportional to the system-size~(see discussion in the main text).}
\end{figure}

Fig.~\ref{fig:TFI_gs} shows the optimization results for the SCom-QAOA circuits for different system sizes as the number of layers is varied for the ground state. The latter occurs in the~$Q = +1$ sector. The top left panel shows the final overlap to the target state,~$O_N(l = N) = |\langle \psi_T|U|\psi_0\rangle|$, at the end of the $N$ applied layers~[see Eq.~\eqref{eq:U} for the definition of~$U$]. For each choice of~$L$, the number of layers required to reach the desired fidelity, denoted by~$N_c$, turned out to be~$L/2$ layers~(top right panel). This is reminiscent of the results for the periodic Ising chain obtained in Ref.~\cite{Ho2019}. However, note that the depth of the SCom-QAOA circuit optimized based on QNG does, in fact, depend on the desired fidelity to the target state and the spectral gap of the model, as described below.

The bottom panels show the behavior of the overlap to the target~$O_N(l)$ and the expectation value of the Hamiltonian~$E_N(l)$ respectively as a function of the layer index~$l$. These two plots show the behavior of the two aforementioned quantities as the system evolves through the different layers of the optimized circuit obtained with~$N = N_c = L/2$. The behavior of~$O_N(l)$ and~$E_N(l)$ are non-monotonic with the non-monotonicity being more severe for larger system-sizes. Intuitively, this can be viewed as a consequence of the fact that the energy levels of a critical chain are split~(to leading order) as~$1/L$. With increasing system-size, the QNG-optimization scheme, which maximizes the overlap to the target state, has a tougher job finding the optimal next iteration step since there are many nearby states. A related phenomenon occurs while searching for a gapped ground state, see Sec.~\ref{sec:Ising_h}. The non-monotonicity of the overlap explains why the number of layers being required to attain desired fidelity scales proportional to~$L$. The quantum circuit evolves the product state with a time-dependent inhomogeneous quench Hamiltonian. We also noticed that in contrast to the linear growth of entanglement entropy between two parts of the system saturating to an extensive value~($\propto L$)~\cite{Calabrese2005, Calabrese2020}, the same grows at first to an extensive value before diminishing to the a logarithmic dependence~($\propto\ln L$) that is characteristic of the critical Ising ground state.  

\begin{figure}\centering
\includegraphics[width = 0.5\textwidth]{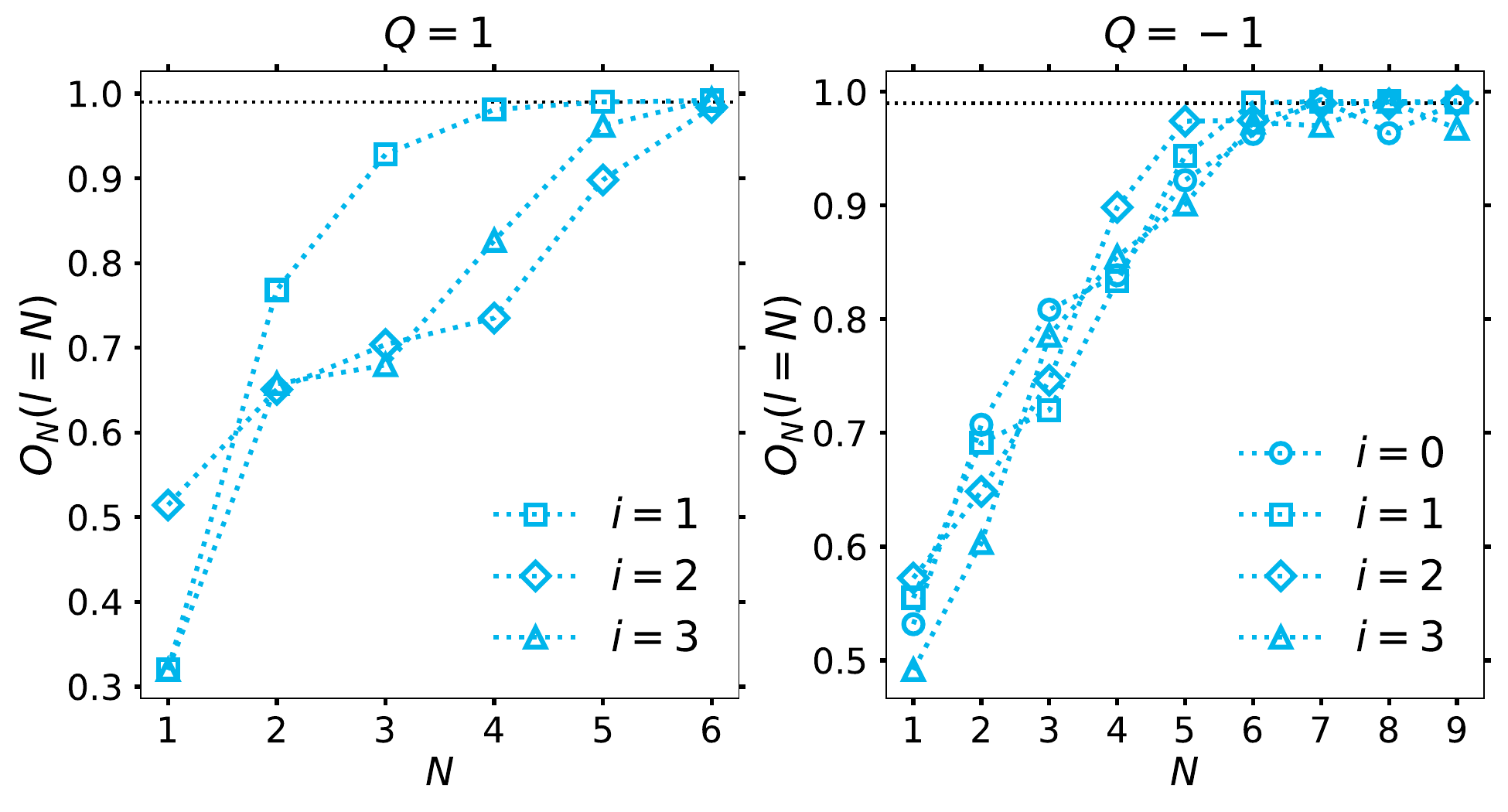}
\caption{\label{fig:TFI_exc} Overlap to the target state  as a function of the number of layers in the~$Q = +1$~(left) and the~$Q = -1$~(right) sectors. The system size was chosen to be $L = 12$. The results are shown for the first seven excited states. The index~$i$ labels the state in a given symmetry sector, with~$Q=1,i=0$ corresponding to the ground state~(see Fig.~\ref{fig:TFI_gs}). The black dotted lines in the two panels  correspond to the desired fidelity of 0.99.}
\end{figure}

Fig.~\ref{fig:TFI_exc} shows the results for the seven lowest excited states for the critical Ising chain. Three of the seven are in the~$Q = +1$ sector, while the remaining are in the~$Q = -1$ sector. By choosing the initial state to be in the suitable symmetry sector, the target states were realized. This is because the relevant symmetry is conserved throughout the quantum circuit evolution. Note that the excited states have higher entropy than the ground state~\cite{Alcaraz2011} and in general, require circuits which are deeper than the one realizing the ground state. For~$L = 12$, this is noticeable for the excited states in the Q = -1 sector~[indexed by $i = 2,3$, Fig.~\ref{fig:TFI_exc}~(right panel)].

\subsection{The Ising chain in longitudinal field}
\label{sec:Ising_h}
Here, the results are shown for the critical Ising chain perturbed by a longitudinal field. The corresponding Hamiltonian is given in Eq.~\eqref{eq:H_I}, with~$\lambda^Z=1,\lambda^X\neq 0$. In the scaling limit, the lattice model is described by the integrable $E_8$ field theory~\cite{Zamolodchikov1989}. However, since our interest is in finding the circuit parameters for transforming a high-energy state to the ground state of the non-integrable lattice model, the integrable characteristic of the continuum model is not relevant. The corresponding layer unitary~$U^l$~[Eq.~\eqref{eq:U_l}] is built out of a layer of $R_{\rm XX}$-s followed by a layer of~$R_{\rm Z}$ and a layer of~$R_{\rm X}$-s~[Fig.~\ref{fig:schematic}(top right)].

\begin{figure}\centering
\includegraphics[width = 0.5\textwidth]{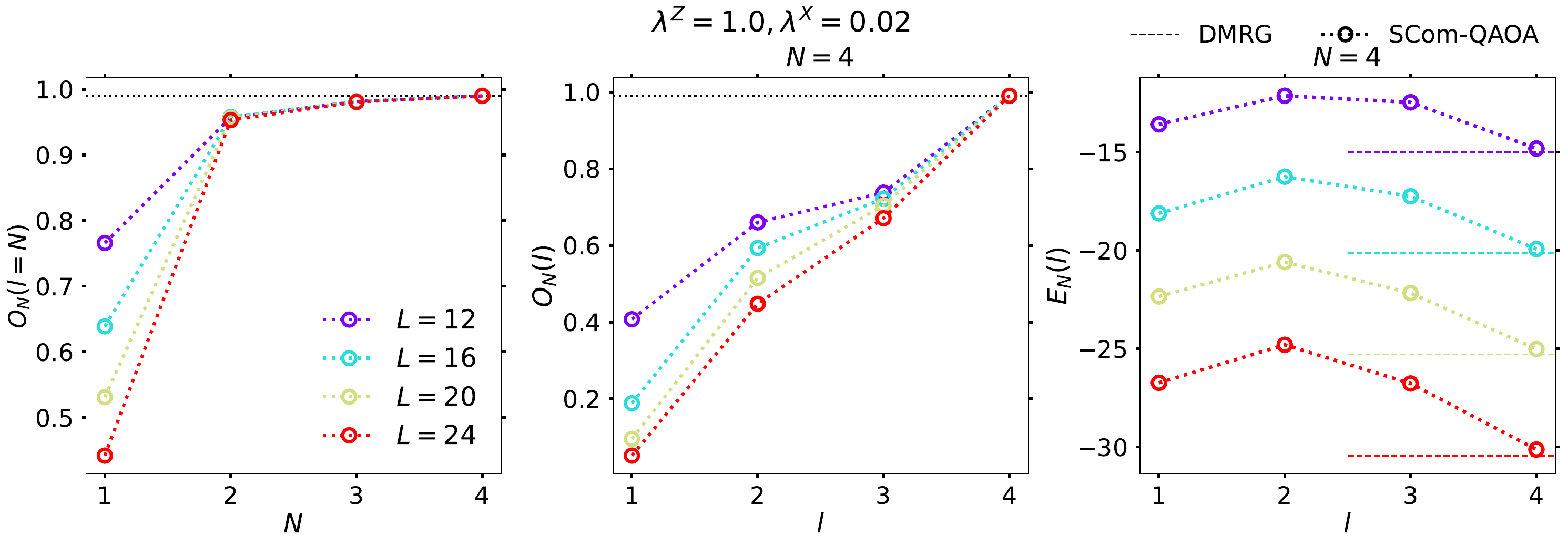}
\caption{\label{fig:TFI_h} Optimization results for the realization of the ground state of the~(gapped) Ising model with longitudinal field model. The parameters chosen were:~$\lambda^Z = 1.0, \lambda^X = 0.06$~[Eq.~\eqref{eq:H_I}]. Variation of the overlap to the target state as a function of the number of layers is shown in the left panel for different system sizes. In contrast to the critical chain, the number of layers~$N_c$ required to reach 99\% fidelity~(black dotted lines in the left and center panels) is independent of the system size.  The variations of the overlap~$O_N(l)$ and the energy~$E_N(l)$ with the number of layers are shown in the center and right panels respectively for~$N = 4$. The variation of the latter three quantities are much smoother and approaches a monotonic behavior as the parameter~$\lambda^X$ increases. This can be understood as a consequence of the spectral gap separating the target ground state from the remaining eigenstates~(see Fig.~\ref{fig:TFI_h_var} and the main text for discussion). We checked that increasing the cut-off for fidelity closer to 1 reduces the discrepancy between the DMRG (shown with dashed line in right panel) and the SCom-QAOA computations.} 
\end{figure}

\begin{figure}\centering
\includegraphics[width = 0.5\textwidth]{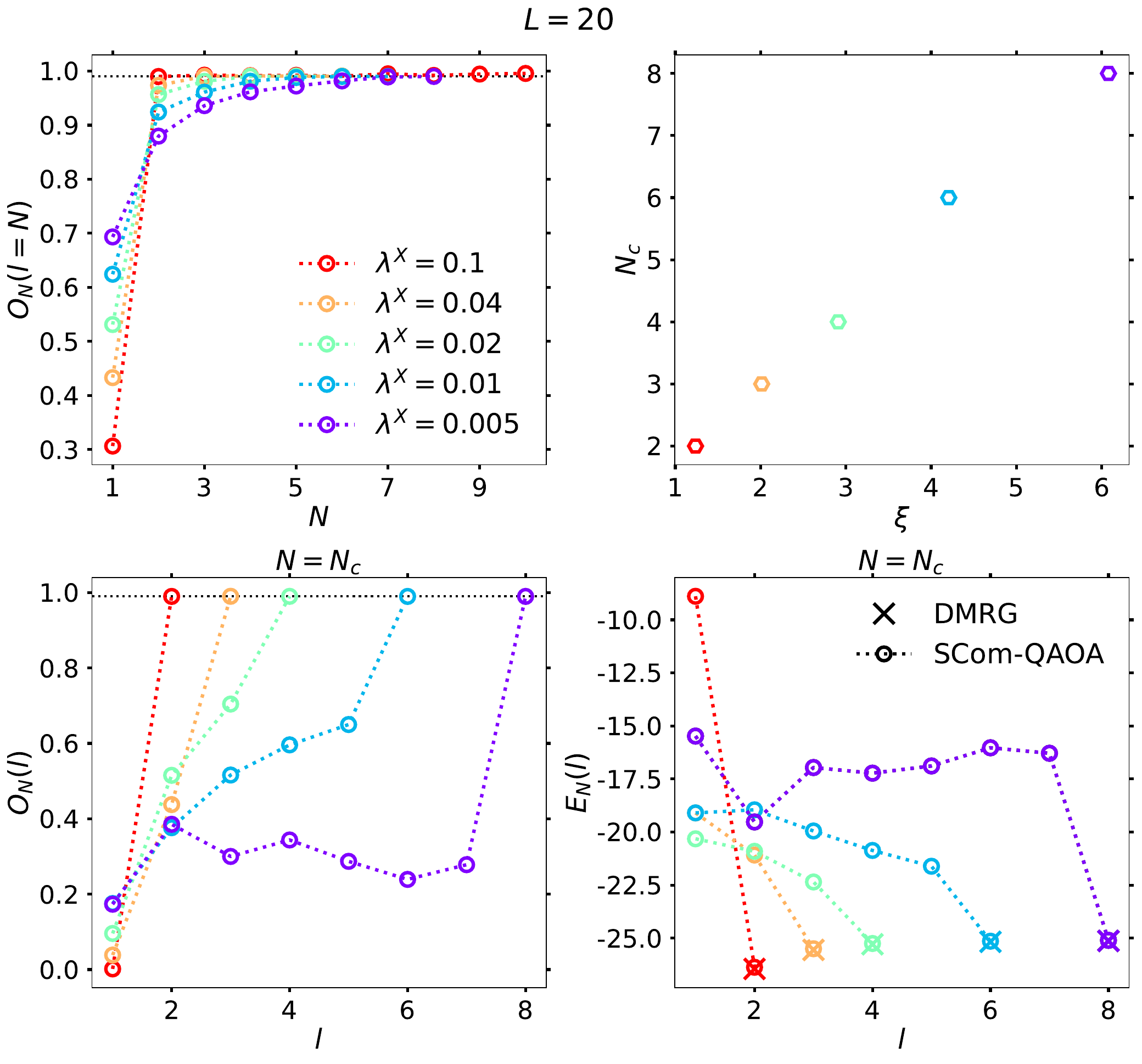}
\caption{\label{fig:TFI_h_var} Optimization results for $L = 20$. (Top left) Variation of the overlap as a function of the number of layers for different choices of~$\lambda^X$~(recall~$\lambda^X = 0$ corresponds to the critical Ising chain). With the increase of~$\lambda^X$, the number of layers required to reach the target state diminishes. (Top right) Variation of the number of layers needed to reach the desired fidelity~$N_c$ with the correlation length~$\xi$ in the ground state is shown. The correlation length plays the role of the effective system size in the gapped case. Its values are obtained using infinite DMRG computations. As seen from the results,~$N_c\propto\xi$. (Bottom panels) Variation of the overlap~$O_N(l)$~(left) and energy~$E_N(l)$~(right) as a function of the layer index~$l$ for the different choices of~$\lambda^X$ when~$N = N_c$. As~$\lambda^X$ increases, the variation approaches a monotonic behavior, which can be attributed to the increased spectral gap~(see main text for discussion).}
\end{figure}

Fig.~\ref{fig:TFI_h} shows the results for generating the ground state of the model for~$\lambda^Z=1.0$,~$\lambda^X=0.06$. In contrast to the critical Ising chain~(Fig.~\ref{fig:TFI_gs}), the number of layers required to reach the desired fidelity to the target state is independent of the system-size~(left panel).  The corresponding overlap~$O_N(l)$ and the energy~$E_N(l)$ are shown in the center and right panels. In contrast to the gapless case, the overlap and energy are monotonic with the layer index~$l$. This is likely a consequence of the spectral gap separating the ground state from the rest of the spectrum. The aforementioned gap enables the QNG-optimization to find a more optimal path towards the target state compared to the gapless case~(see also Fig.~\ref{fig:TFI_h_var}). Note that for small system-sizes, the variations are `less monotonic'. This is to be expected since the true gapped behavior is only apparent when the system-sizes are large enough compared to the correlation length~(see top right panel of Fig.~\ref{fig:TFI_h_var} for values for the latter).

Fig.~\ref{fig:TFI_h_var} shows the variation of the required number of layers with the parameter~$\lambda^X$. As the latter is decreased, the energy gap between the ground and the first excited states diminishes~(recall that~$\lambda^X = 0$ is the critical point). This leads to an increase in the required number of layers to reach the desired fidelity to the target state. The top panels show the overlaps~$O_N(l= N)$ and the number of layers,~$N_c$, required to reach the desired fidelity. The latter quantity~(top right panel) is plotted as a function of the correlation length~$\xi$, which is obtained using infinite DMRG computations. The top right panel suggests a dependence~$N_c\propto\xi$, which is reminiscent of the gapless case result with the system-size replaced by the correlation length. The overlaps~$O_N(l)$ and the energies~$E_N(l)$ are shown in the bottom panels as a function of the layer index~$l$ for $N = N_c$. As explained earlier, for smaller value of~$\lambda^X$ which is associated with a smaller spectral gap, the~$O_N(l),E_N(l)$ start deviating the monotonic behavior. 

\begin{figure}\centering
\includegraphics[width = 0.5\textwidth]{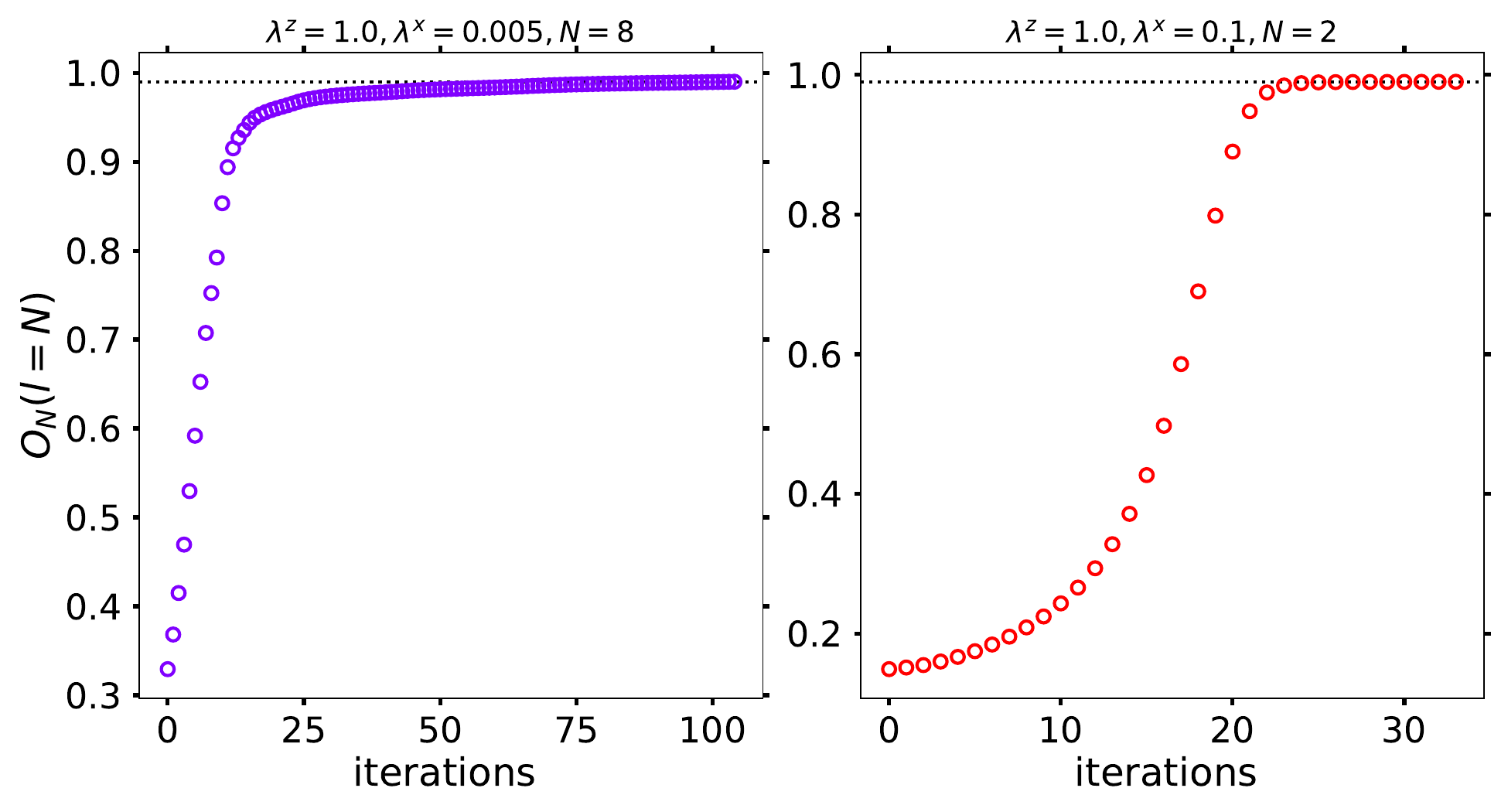}
\caption{\label{fig:ovs_iter} Variation of the overlap as a function of the iteration steps for $L = 20$. The corresponding model parameters~[Eq.~\eqref{eq:H_I}] are shown in the panel titles. As is evident from the two curves, the overlap to the target state grows monotonically with the iteration steps. The black dotted lines in the two panels correspond the target fidelity of 0.99. }
\end{figure}

Before concluding we note that the number of iterations required for achieving the target fidelity is less than 100. Fig.~\ref{fig:ovs_iter} shows the results for the gapped Ising model~[Eq.~\eqref{eq:H_I}] with~$L = 20$ for two different choices of parameters. Results for the other models are similar and not shown for brevity. 

\subsection{The Tricritical Ising model}
\label{sec:TCI}
Next, results are presented for the ground state of a perturbed Ising model, which for certain choice of parameters, realizes the tricritical Ising model~\cite{Brien2018}. The Hamiltonian is given by
\begin{align}
\label{eq:H_TCI}
H^{\rm TCI} &= - \lambda^Z\sum_{j = 1}^LZ_j -\sum_{j = 1}^{L-1}X_{j}X_{j+1}\\\label{eq:H_nnn}\qquad &+\lambda^{\rm ZXX}\sum_{j = 1}^{L-2}\big(X_jX_{j+1}Z_{j+2} + Z_jX_{j+1}X_{j+2}\big).
\end{align}
While for~$\lambda^{\rm ZXX} = 0$,~$H^{\rm TCI}$ reduces to the Hamiltonian of the ordinary transverse-field Ising model, for~$\lambda^{\rm ZXX} \approx 0.428, \lambda^Z = 1.0$, the long-wavelength properties of this model is described by the tricritical Ising model~\cite{Brien2018}. Below, the SCom-QAOA circuits for the realization of the tricritical Ising ground state is described. 

Following the scheme described in Sec.~\ref{sec:QNG}, each layer unitary~$U^l$ is built out of a sublayer of~$R_{\rm XX}$, followed by a sublayer of~$R_Z$ and subsequently, a layer of the three-body interacting term~$R_{\rm ZXX}$~[see Fig.~\ref{fig:schematic}(e)]. Note that the different terms in the summation of Eq.~\eqref{eq:H_nnn} do not commute amongst themselves. The second-order Suzuki-Trotter decomposition was used for the sublayer of~$R_{\rm ZXX}$ operators~\footnote{We verified that using the first order Trotter decomposition led to only quantitative changes in the results.}. As explained in Sec.~\ref{sec:QNG}, the angles for the different unitary operators constituting~$U^l$ are site-dependent. In an actual quantum simulator, the three-body rotation would be further decomposed into one and two-qubit gates using standard techniques~(see, for example, Chap. 4 of Ref.~\cite{Nielsen_Chuang_2000}). The TEBD computations were done with three-site unitary operator. Notice that the quantum circuit evolution conserves the~$\mathbb{Z}_2$ symmetry of~$H^{\rm TCI}$ associated with the operator is~$\prod_jZ_j$. 

\begin{figure}\centering
\includegraphics[width = 0.5\textwidth]{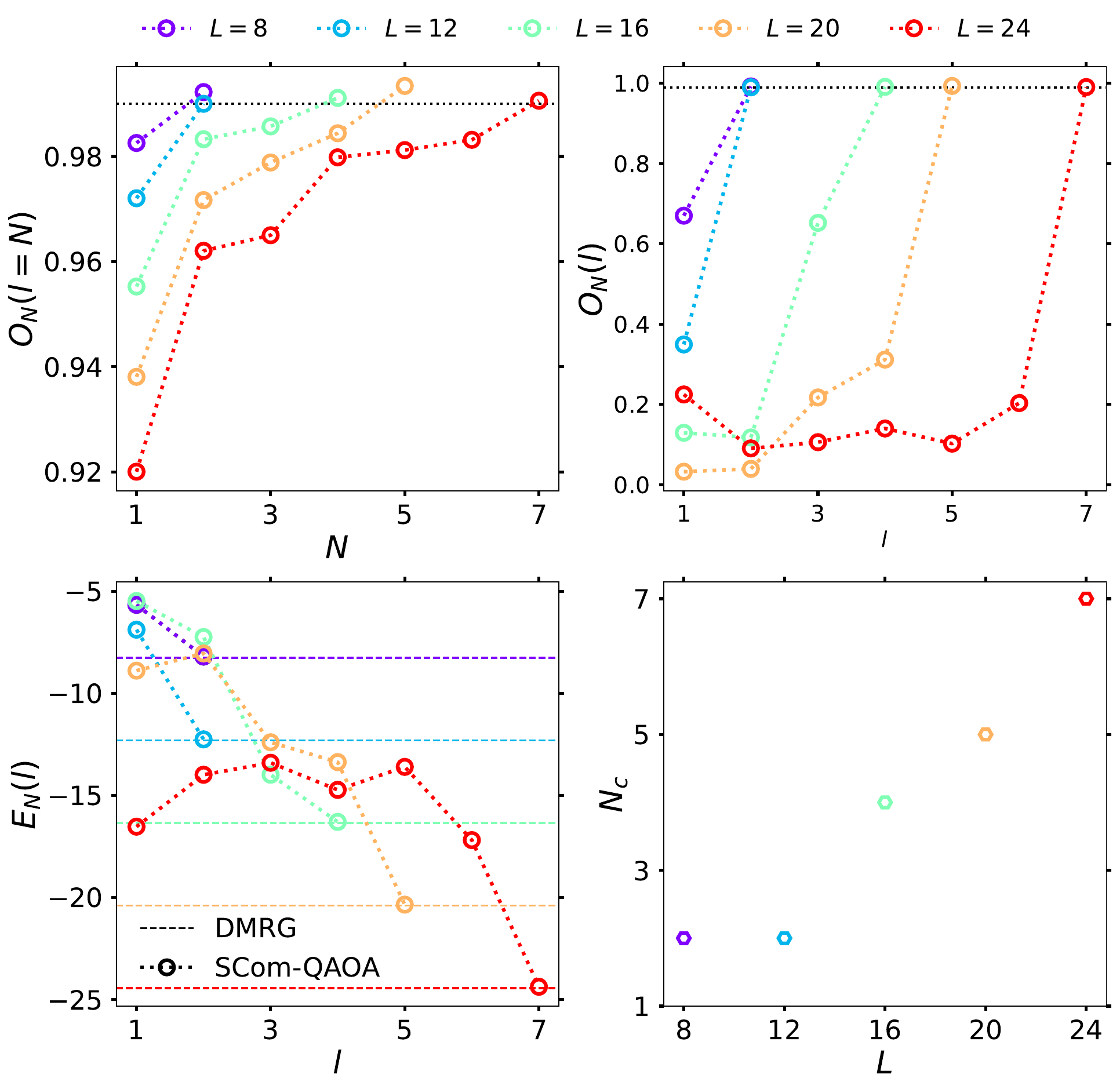}
\caption{\label{fig:TCI} Optimization results for the ground state of~$H^{\rm TCI}$ at the tricritical Ising point~[Eq.~\eqref{eq:H_TCI}].~(Top left) Variation of the overlap to the target state as a function of the number of layers. Note that the desired fidelity is reached faster~($N_c\sim L/4$, see bottom right panel) than the Ising case~(see Fig.~\ref{fig:TFI_gs}). Intuitively, this can be understood as a consequence of the rapid build-up of entanglement resulting from the three-site unitary operations~$R_{\rm ZXX}$. The top right and the bottom left panels show the variation of the overlap~$O_N(l)$ and  energy~$E_N(l)$ with the layer index $l$. In these two plots,~$N$ is chosen to be the number of layers which is sufficient to reach the desired fidelity. The variation of $O_N(l), E_N(l)$ remain non-monotonic as was true for the critical Ising chain~(see Fig.~\ref{fig:TFI_gs} and discussion in Sec.~\ref{sec:crit_Ising}).}
\end{figure}

Fig.~\ref{fig:TCI} shows the results for the realization of the ground state of~$H^{\rm TCI}$ for~$\lambda^Z = 1, \lambda^{\rm ZXX} = 0.428$. Notice again the growth of the number of required layers to attain the 99\% fidelity cut-off with system-size. However, the growth is slower than the Ising case~(bottom right panel) with the desired fidelity achieved in~$\sim L/4$ layers. Intuitively, this can be understood as a consequence of the rapid build-up of entanglement due to the three-qubit gates instead of the Ising case which had only two-qubit gates. It is possible that, in general, for critical models, the SCom-QAOA circuits lead to~$N_c\propto\beta L$. Here,~$\beta$ is a function of not only the desired fidelity and the critical model being investigated, but also other non-universal factors and we leave it as a future problem to find out the full functional dependence. 
The variation of the overlap~$O_N(l)$ and the energy~$E_N(l)$  with the layer index~$l$ are shown in the top right and the bottom left panels. Their dependence remain non-monotonic, as for the critical Ising chain with the non-monotonicity more severe as the system-size increases~(see Sec.~\ref{sec:crit_Ising} for discussion). The lowest few excited states in the~$Q = \pm 1$ sectors were obtained. The results are qualitatively similar as for the critical Ising case and are not shown for brevity. 

\section{Summary and Outlook}
\label{sec:concl}
To summarize, this work proposes a parametrized quantum circuit ansatz for efficient preparation of arbitrary entangled eigenstates of quantum many body Hamiltonians. The proposed SCom-QAOA circuits are a symmetry-conserving modification of the QAOA operator ansatz. The parameters of the proposed quantum circuit are determined using the steepest gradient descent algorithm based on the Quantum Natural Gradient. The latter takes into account the curvature of the manifold of quantum states while performing gradient descent. Explicit demonstrations are provided for the critical Ising chain, the Ising model in a longitudinal field and the tricritical Ising model. The circuit depth was found to scale linearly with the length of the chain for ground states of critical models. The same was found to be independent of the system-size and determined entirely by the finite correlation length for gapped ground states. While the results were presented for desired fidelity of 99\%, the SCom-QAOA schemes allow efficient preparation of higher-fidelity states as well. This was verified explicitly for a fidelity cut-off of 99.9\%, which led to only quantitative changes in the results. Finally, the SCom-QAOA circuits can be straightforwardly generalized to target different states without conserving any symmetry. While this did not lead to a qualitative change of the scaling of the circuit depth for the critical Ising model, this can happen for other models~\footnote{We thank Frank Pollmann for bringing this point to our attention.}. 

The proposed efficient preparation of entangled many-body states could be useful for a wide range of practical quantum algorithms. First, it could be used to probe a wider family of quench dynamics than what has been done on digital quantum simulators~\cite{Bernien_2017, Smith2019, Vovrosh2021, Lamb2023}. Second, an approximate ground state computed using classical methods could be a starting point for a more accurate determination of the actual ground state using hybrid quantum algorithms such as the variational quantum eigensolver~\cite{Peruzzo2011}, the variational quantum simulator~\cite{Li2017} and the eigenstate witnessing approach~\cite{Santagati2018}. 

Before concluding, we outline two future research directions. First, the current work used the overlap cost function~[Eq.~\eqref{eq:L_O}] to determine the circuit parameters. Naturally, this assumes that the target state is known from other techniques. Generalizations to the case when the target state is unknown is straightforward using the Hamiltonian cost function~[Eq.~\eqref{eq:L_H}]. This would be particularly useful for models which lie beyond the reach of conventional numerical techniques like density-matrix renormalization group. A digital quantum simulator, coupled to classical optimization routines, could be used to investigate these models. In this case, the computation of the {\it entire} Fubini-Study metric as well as the gradients of the cost-functions would have to be done using measurements on digital quantum simulators. Some schemes have been proposed to achieve this goal~\cite{Gacon2021}. Note that the SCom-QAOA circuits can be readily implemented for the investigation of two-dimensional quantum Hamiltonians, critical or otherwise. We will report our findings on this topic in a separate work.  

Second, the QNG-based optimization problem of the SCom-QAOA circuits can be viewed as a more restrictive case of the general problem of determining circuit complexity of quantum many-body Hamiltonian ground states~\cite{Nielsen2005, Nielsen2006, Dowling2006}. A more general solution, which would enable determining the actual circuit complexity, would involving optimization in space of all possible unitary rotations and minimizing the distance between the desired unitary rotation and the identity~\cite{Nielsen2005}. While the QNG-based optimization finds an optimal path for the specific choice of the SCom-QAOA ansatz with the given cost-function, it does not rule out `a more direct' path to the target state. A more quantitative investigation of generalizations of the SCom-QAOA circuits and comparison with Nielsen's bound could potentially help construct the optimal circuits for given target eigenstates of generic strongly-interacting Hamiltonians~\cite{DiGiulio:2020hlz, Jefferson:2017sdb, Craps2023}.

\section*{Acknowledgements}
Discussions with Sergei Lukyanov, Frank Pollmann, and Yicheng Tang are gratefully acknowledged. This work was supported by the U.S. Department of Energy, Office of Basic Energy Sciences, under Contract No. DE-SC0012704. 

\bibliography{/Users/ananda/Dropbox/Bibliography/library_1}
\end{document}